\documentclass[letterpaper, 10 pt, conference]{ieeeconf}  

\IEEEoverridecommandlockouts                              

\usepackage{amsmath}
\usepackage{cleveref}
\usepackage[utf8]{inputenc}
\usepackage[english]{babel}
\usepackage[sorting = none]{biblatex}
\usepackage{lipsum}
\usepackage{mwe}

\usepackage{longtable}
\usepackage{url}

\usepackage{graphicx} 
\graphicspath{ {./images/} }
\usepackage{booktabs}
\usepackage[load-configurations=version-1]{siunitx} 
\usepackage{mathtools}
\usepackage{amsmath}
\usepackage{ragged2e}
\usepackage{array}
\usepackage[super]{nth}
\usepackage{csquotes}

\begin{document}

\title{\LARGE \bf IMLE-Net: An Interpretable Multi-level Multi-channel Model for ECG Classification}

\author{Likith Reddy  \and
Vivek Talwar \and
Shanmukh Alle \and
Raju. S. Bapi \and
U. Deva Priyakumar
}

\author{Likith Reddy$^{1}$, Vivek Talwar$^{1}$, Shanmukh Alle$^{2}$, Raju. S. Bapi$^{3}$, U. Deva Priyakumar$^{2}$\\

	\normalsize $^{1}$International Institute of Information Technology, Healthcare and AI Center, Hyderabad\\
	\normalsize $^{2}$International Institute of Information Technology, CCNSB Lab, Hyderabad\\
	\normalsize $^{3}$International Institute of Information Technology, Cognitive Science Lab, Hyderabad\\
	\textit{Email: likith.reddy@ihub-data.iiit.ac.in}
}










\maketitle

\begin{abstract}

Early detection of cardiovascular diseases is crucial for effective treatment and an electrocardiogram (ECG) is pivotal for diagnosis. The accuracy of Deep Learning based methods for ECG signal classification has progressed in recent years to reach cardiologist-level performance. In clinical settings, a cardiologist makes a diagnosis based on the standard 12-channel ECG recording. Automatic analysis of ECG recordings from a multiple-channel perspective has not been given enough attention, so it is essential to analyze an ECG recording from a multiple-channel perspective. We propose a model that leverages the multiple-channel information available in the standard 12-channel ECG recordings and learns patterns at the beat, rhythm, and channel level. The experimental results show that our model achieved a macro-averaged ROC-AUC score of 0.9216, mean accuracy of 88.85\% and a maximum F1 score of 0.8057 on the PTB-XL dataset. The attention visualization results from the interpretable model are compared against the cardiologist's guidelines to validate the correctness and usability.

\end{abstract}


\section{Introduction}
\justify
Heart diseases are the leading cause of death globally. It is estimated that 17.9 million people died of heart diseases in the year 2016 which represents about 31\% of global human mortality according to the World Health Organization. Early detection of heart disease is crucial for effective treatment and reducing mortality. Electrocardiography is an inexpensive, simple and non-invasive procedure that helps us to understand the functioning of a heart which in turn helps us diagnose heart diseases. An Electrocardiogram (ECG) records information about the electrical activity of the heart based on which a cardiologist can identify abnormal functioning of the heart to diagnose various conditions. But this process of analyzing an ECG recording is time-consuming that requires a trained expert’s time and attention. Moreover, it is prone to human error. Misdiagnosis of early signs of heart diseases by cardiologists from an ECG recording is still a major concern~\cite{12345}. So work has been done to automate the ECG signal classification to aid in human diagnosis.

Most existing approaches use traditional machine learning-based methods for the detection of heart diseases from a patient’s ECG recording~\cite{7519646, mlapproach}. Such methods involve manually crafted feature extraction and the use of classifiers on the extracted features. Asgari et al.~\cite{asgari2015automatic} use the stationary wavelet transform to extract features and a support vector machine (SVM) for the detection of atrial fibrillation. Poddar et al.~\cite{poddar} employed the heart rate variability (HRV) features computed in time-domain, frequency-domain and using non-linear methods, the features are then classified using k-nearest neighbor (KNN) and SVM classifiers. Unsupervised methods using time series clustering were proposed in Annam et al.~\cite{Annam2016}.

The advent of deep learning-based methods for ECG signal classification has progressed in recent years to reach cardiologist-level performance~\cite{rajpurkar}. Such methods analyze ECG data without manual feature extraction in contrast to traditional machine learning approaches. The work done by Rajpurkar et al.~\cite{rajpurkar} uses a time series ECG signal as an input to a 34-layer  eural network (CNN) by mapping a sequence of ECG samples to a sequence of rhythm classes. The model exceeded the average cardiologist performance in both precision and recall. Mousavi et al.~\cite{8683140} developed an automatic heartbeat classification method using deep convolutional neural networks and sequence-to-sequence models. Murugesan et al.~\cite{8438739} proposed a combination of CNNs and a long short-term memory (LSTM) based feature extractor that can be directly trained without any preprocessing. Recurrent neural network (RNN) along with an attention mechanism is used by Schwab et al.~\cite{8331750} for the classification of a single-channel ECG recording.

An interpretable model along with reliable performance is crucial to instil confidence in medical practitioners to use computer-assisted electrocardiography. A lot of work has been done to improve the interpretability of deep learning models in computer vision tasks but the progress made on interpreting ECG classification models has been limited and has been gaining interest recently~\cite{8834637, yong}. Vijayarangan et al.~\cite{9176396} use a gradient-weighted class activation map~\cite{8237336} for visualizing saliency on single-channel ECG signals using their proposed model. Attention models have been utilized by Hong  et  al.~\cite{ijcai2019-816} to improve interpretability in the time and frequency domains.

The majority of the interpretable deep learning work focuses on models which deal with single-channel ECG recordings. A single-channel ECG recording is used for basic heart monitoring whereas a multi-channel ECG recording provides information about the 3-dimensional electrical activity of the heart. From a clinical point of view, a cardiologist makes a diagnosis based on the standard 12-channel ECG recording~\cite{12lead}. When compared to a single-channel recording, a 12-channel recording accurately reproduces various features of an ECG such as QRS complex, ST-segment and T waveforms which sometimes are poorly represented in a single-channel recording~\cite{kligfield2001value}. For example, ST-segment elevation which is an important characteristic for Myocardial Infarction (MI) is best identified using a 12-channel recording and the localization of myocardial injury cannot be identified without a 12-channel recording. Analysis of ECG recordings from a multiple-channel perspective has not been given enough attention and it is essential to analyze an ECG recording from a multi-channel perspective. To address this gap, we propose a model that leverages the multi-channel information available in a multi-channel ECG recording. Also, we make use of the hierarchical structure of the ECG signal to learn patterns at the beat, rhythm and channel level. The model is then evaluated on the PTB-XL dataset~\cite{Wagner:2020PTBXL} which comprises of a total 21837 clinical 12-channel ECG recordings of 10 seconds length from 18885 patients. The performance of the model is then compared against the existing deep learning methods. The interpretability of our model is evaluated by comparing the attention visualizations with the cardiologist's rules for identifying the subtypes of MI. We perform an ablation study by reducing the number of available channels in an input ECG recording to evaluate the robustness of the proposed model. The source code of the proposed model and all the experiments performed is made publicly available to ensure reproducibility for future research\footnote[1]{\url{https://github.com/likith012/IMLE-Net}}.

\section{Methodology}

\subsection{Preprocessing}
In preprocessing, a multi-channel ECG signal is segmented into beat segments of length $W$ for every individual channel. A popular approach to segment an ECG signal into beat segments is based on first detecting the R-peaks in an ECG signal using the Pan-Tompkins algorithm~\cite{4122029} and then using the adaptive searching window for segmenting beats which is used in Mousavi et al.~\cite{mousavi2020han}. Other approaches for the detection of R-peaks include standard wavelet transform~\cite{MERAH2015149} and Hamilton~\cite{1166717}. But these approaches perform poorly in the presence of signal noise and HRV irregularities. Instead, we opted for a sliding window approach with no overlap which was used by Hong et al.~\cite{ijcai2019-816} to segment beats of the ECG signal. To get the $k$\textsuperscript{th} beat segment, the beat is spanned from $(k -1) \times W$ to $k \times W$ over the ECG signal, where $W$ is the length of the window.

\subsection{Architecture}
\begin{figure}[htbp]
    \centering
    \includegraphics[width=3in]{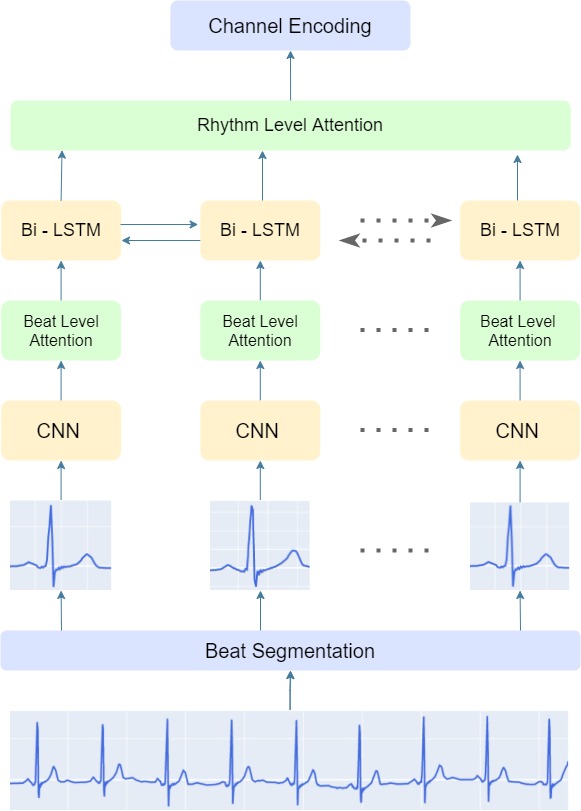}
    \caption{Architecture of the model at each individual channel}
    \label{Fig:Fig1}
\end{figure}

\begin{figure}[htbp]
    \centering
    \includegraphics[width=3in]{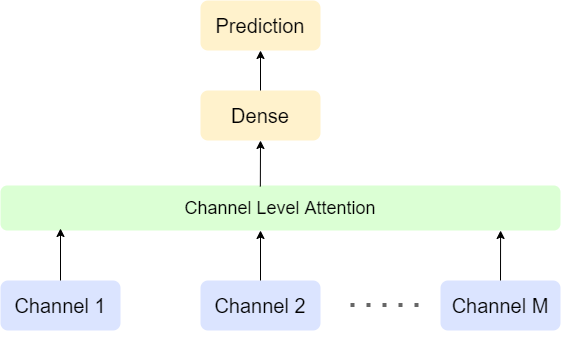}
    \caption{Channel level architecture of the model}
    \label{Fig:Fig2}
\end{figure}

The proposed model can be divided into two parts, the first part processes each channel separately to generate a channel encoding for an input ECG recording. The architecture of the first part is shown in~\Cref{Fig:Fig1}. The segmented beats from a multi-channel ECG recording are passed through a beat level block, the outputs from the beat level block are passed through a rhythm level block to get channel-wise encodings. These channel-wise encodings are passed into the second part of the model which pools these encodings from each channel and generates predictions as shown in~\Cref{Fig:Fig2}.

\subsubsection{Beat Level Block}
The beat level block involves the processing of input beat segments. The beats are first passed through a CNN and then an attention layer.

\begin{figure}[htbp]
    \centering
    \includegraphics[width=2in]{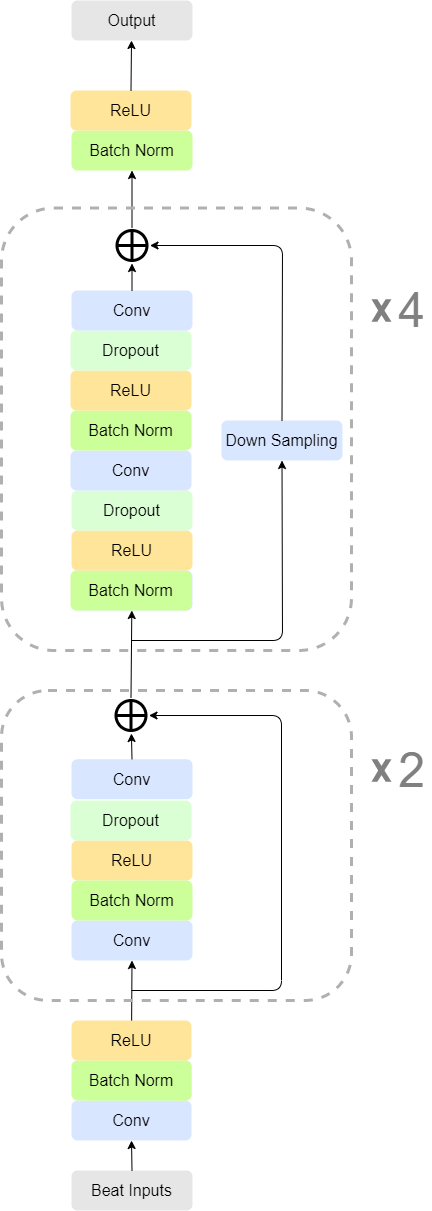}
    \caption{Architecture of the CNN used at the beat level.}
    \label{Fig:Fig3}
\end{figure}

The architecture of the CNN used in the beat level block is shown in \Cref{Fig:Fig3}. Inspired by the ResNet architecture~\cite{He_2016_CVPR}, the CNN uses a total of 6 residual blocks. Skip connections in a residual block improve backpropagation in deep neural networks which optimize model training. In a single residual block, there are two convolutional layers and every convolutional layer is followed by a batch normalization layer and a ReLU activation function. Between two convolutional layers, we employ dropout to stabilise model training and to avoid overfitting. No downsampling is done in the first two residual blocks. In the remaining residual blocks, inputs are downsampled by a factor of two after every alternate residual block. All the convolutional layers have a filter length of 8 with 32 starting filters. In every residual block where downsampling of inputs occurs, the number of filters is doubled. All the weights in a CNN are shared across all the beat segments and across multiple channels in the model.

Let $T$ be the length of an ECG signal for the $c$\textsuperscript{th} channel and $W$ be the beat length. The total number of beat segments are given by $N=\frac{T}{W}$. Let $(b_{1}^{c}, b_{2}^{c}, b_{3}^{c}, ... b_{N}^{c})$ be the beat segments for the $c$\textsuperscript{th}  channel of an ECG recording. The CNN takes in a time series of $k$\textsuperscript{th} beat segment $b_{k}^{c}$ of length $W$. The  output from the CNN with input as $b_{k}^{c}$ beat segment be $\bar{b}_{k}^{c}$ with an intermediate sequence length of $\bar{W}$.

We know that all the areas within a heartbeat segment are not equally relevant in identifying an abnormal heartbeat. The attention mechanism helps in identifying more important areas within a beat segment by giving higher attention scores to them. It is a two-layer neural network that takes inputs from the CNN outputs and gets the attention scores by computing softmax over the outputs from the two-layer neural network as given in \cref{eq1:label1,eq2:label2}. The weights of the beat level attention layer are shared across all beat segments similar to CNN.

\begin{equation}
\label{eq1:label1}
{\centering\alpha_{k}^{c} = softmax(V_{b} \tanh({W_{b} \bar{b}_{k}^{c} + b_{b}}))}
\end{equation}
\begin{equation}
\label{eq2:label2}
\centering    {B_{k}^{c} = \sum_{i=1}^{\bar{W}}\alpha_{k, i}^{c}\bar{b}_{k, i}^{c}}
\end{equation}

Where, $V_{b}$, $W_{b}$ and $b_{b}$ are beat level attention parameters that are learned during training. $\alpha_{k}^{c}$ is the beat level attention score for the $k$\textsuperscript{th} beat segment in an ECG signal for the $c$\textsuperscript{th} channel. $B_{k}^{c}$  is the beat context vector for the $k$\textsuperscript{th} beat segment in an ECG signal for the $c$\textsuperscript{th} channel. The beat context vector $B_{k}^{c}$  is a weighted sum of $\bar{b}_{k, i}^{c}$ with weights given by attention scores $\alpha_{k, i}^{c}$.

\subsubsection{Rhythm Level Block}

The rhythm level block deals with the input ECG rhythm signal which is constructed from several beat segments. It consists of two layers, a Bi-directional LSTM and a rhythm level attention layer. Bi-directional LSTM is used as the current state can process both forward and backward directions of the input sequence information. The input to the Bi-LSTM is a sequence of beat context vectors $B_{k}^{c}$  for the $c$\textsuperscript{th} ECG channel with $N$ beat sequences. In the Bi-LSTM, the forward and the backward direction hidden states for the current step $k$ are concatenated to obtain output $r_{k}^{c}$.

The output from the Bi-LSTM  $r^{c}$  with a sequence length of $N$ for the $c$\textsuperscript{th} channel is passed through a rhythm attention layer. The rhythm attention mechanism used here identifies the important beat segments within an ECG rhythm by giving higher attention scores to them and is given by \cref{eq3:label3,eq4:label4}. 

\begin{equation}
\label{eq3:label3}
{\centering\beta^{c} = softmax(V_{r} \tanh({W_{r} r^{c} + b_{r}}))}
\end{equation}
\begin{equation}
\label{eq4:label4}
    {R^{c} = \sum_{i=1}^{N}\beta_{i}^{c} r_{i}^{c}}
\end{equation}

Where, $V_{r}$, $W_{r}$ and $b_{r}$ are rhythm level attention parameters that are learned through training and $\beta^{c}$  is the rhythm level attention score for the $c$\textsuperscript{th} channel. $R^{c}$  is the rhythm context vector for the $c$\textsuperscript{th} channel which encodes entire ECG signal for a particular channel into a vector (Channel Encoding).
\subsubsection{Channel Level Block}
The channel level block processes the information present in several ECG channels. The information present in individual ECG channels differ. The characteristic information for diagnosing a particular condition might be present only in few channels. So we need a mechanism to not treat each channel equally always and give more importance to the channels that hold the characteristics for a particular heart disease. So a channel-level attention mechanism is used here which receives inputs $(R^{1}, R^{2}, ... R^{c} ...  R^{M})$, where $M$ is the total number of channels available in an ECG recording. The channel level attention scores are calculated similar to beat and rhythm level attention given by \cref{eq5:label5,eg6:label6}.

\begin{equation}\label{eq5:label5}
\centering{\gamma = softmax(V_{c} \tanh(W_{c}R^{c} + b_{c}})
\end{equation}
\begin{equation}\label{eg6:label6}
\centering{C=\sum_{i=1}^{M} \gamma_{i}R_{i}^{c}}   
\end{equation}

Where, $V_{c}$, $W_{c}$ and $b_{c}$ are the channel level parameters that are learned during training. $\gamma$ is the channel level attention score and $C$ is a context vector that encodes the entire ECG signal for all the available channels. The context vector $C$ is passed through a linear neural network for final predictions.


\section{Experiments}

\subsection{Dataset}
The dataset used is the PTB-XL dataset~\cite{Wagner:2020PTBXL} which is the largest openly available dataset that provides clinical 12 channel ECG waveforms. It comprises 21837 ECG records from 18885 patients of 10 seconds length which follow the standard set of channels (I, II, III, aVL, aVR, aVF, V1–V6). The dataset is balanced concerning sex with 52\% male and 48\% female and covers age ranging from 0 to 95 years. The dataset covers a wide range of pathologies with many different co-occurring diseases. The ECG waveform records are annotated by two certified cardiologists. Each ECG record has labels assigned out of a set of 71 different statements conforming to the Standard communications protocol for computer assisted electrocardiography (SCP-ECG) standard. The ECG waveform was originally recorded at a sampling rate of 400 Hz and downsampled to 100 Hz. All the experiments in our work were performed using the 100 Hz sampling rate.

\subsection{Evaluation}
We select the samples from the PTB-XL dataset~\cite{Wagner:2020PTBXL} in a way to cover a huge cohort of diseases and use it to evaluate the model. Each SCP statement annotated in the ECG recording belongs to one of the groups of form, rhythm, and diagnostic statement groups, with a few SCP statement annotations belonging to both rhythm and diagnostic statements. For evaluating the model in our work only the diagnostic statements are considered, a total of 21430 ECG recordings are present in the dataset with at least one diagnostic statement. A specific ECG recording can have multiple diagnostic statements annotated against it. For the diagnostic statement labels, hierarchical groups are provided in terms of 5 superclasses and 24 subclasses. In our experiments, we have used the 5 superclass labels in which the ECG recordings in a particular superclass belong to the same family of pathologies. The superclass disease groups are Normal ECG (NORM), Conduction Disturbance (CD), Myocardial Infarction (MI), Hypertrophy (HYP), and ST/T changes (STTC). Splitting of the dataset for training and testing is based on the stratified folds provided by the dataset. Each ECG recording is assigned to one of the ten folds, folds 1 to 8 are used for training the model, the 9\textsuperscript{th} fold is used as the validation set and the 10\textsuperscript{th} fold is used as a test set. The train and test sets are standardized before training.

The proposed model is compared against Resnet 101 model~\cite{He_2016_CVPR}, Mousavi et al.~\cite{8683140}, ECGNet~\cite{8438739} and Rajpurkar et al.~\cite{rajpurkar}. The performance of the model is evaluated using metrics such as the mean accuracy, macro averaged Area under Receiver Operating Characteristics (ROC-AUC) and maximum F1 score. Along with these metrics, class-wise accuracy and class-wise ROC-AUC  scores are also reported.

\subsection{Interpretability }
We used the beat attention scores, rhythm attention scores and channel attention scores to gain insights into the model at beat, rhythm and channel level.

As the training samples prepared in evaluating the model are grouped according to the same family of pathologies, such training data is not ideal to validate the channel level interpretations as the grouped ECG recordings do not hold many variations in abnormal beat characteristics across different ECG channels. For validating channel level interpretations, we chose two specific sub-diagnostic diseases of MI, namely, Anteroseptal Myocardial Infarction (ASMI) and Inferior Myocardial Infarction (IMI) along with normal (NORM) ECG recordings and rejected the rest of the subtypes of MI mainly due to the paucity of the number of ECG recordings for these subtypes. A total of 1344 (ASMI), 916 (IMI), and 1000 (NORM) samples are obtained. Normal ECG samples are obtained by randomly selecting 100 samples from each fold. We have chosen the stratified folds from 1 to 8 as training data and folds 9 and 10 as test data.

\subsection{Ablation Study}
We experimented our model with different combinations of ECG recording channels (leads), by doing so we can estimate how well these combinations of channels contribute to the classification of superclass labeled data which was used in the evaluation of the model. The channel combinations selected are limb leads (I,  II,  III,  aVL,  aVR and  aVF), precordial leads (V1-V6), leads I, II and II and all the available 12 leads in the ECG recording. The performance of our proposed model is compared against different combinations of ECG channels and class-wise ROC-AUC and class-wise accuracy scores are reported.

\subsection{Implementation Details} 

The model is trained for a maximum of 60 epochs in mini-batches of 32 with an early stopping criterion. The Cross-Entropy loss function has been used for training the model with Adam optimizer~\cite{kingma2017adam} with the first momentum value set to 0.9 and the second momentum value set to 0.99. An adaptive learning rate is used for minimizing the loss with an initial learning rate set to 0.001. The learning rate is decreased by a factor of 10 after every 10 epochs if no improvement in performance is observed.  Training is done with window size ($W$) of 50 time-points and a dropout rate of 0.5. An L2 regularisation loss with a coefficient of ($2 \times 1e$-5) has been used to avoid overfitting of the model. The model is developed using Tensorflow and all the experiments are trained on an Nvidia RTX 2080Ti 11GB GPU workstation.

\section{Results}
A comparison of class-wise ROC-AUC and class-wise accuracy scores across the models is shown in \Cref{table:1,table:2}, respectively. Our model achieves the best class-wise ROC-AUC score across all the classes except for the HYP class and the class-wise accuracy score is highest for all classes except for the CD class. Table \ref{table:3} shows that our model achieved the best performance across the metrics macro ROC-AUC, mean accuracy and maximum F1 score as compared to the other models. 
This is due to the effective use of the multi-channel information used by our model compared to other models by incorporating a channel level attention. The processing of the hierarchical structure of an ECG signal is efficiently used by using beat and rhythm level attention. The model parameters are shared across multiple channels instead of having separate model parameters for each channel resulting in fewer final model parameters and better generalization.

\begin{figure*}[htbp] 

  \centering
  \includegraphics[width=0.58\textwidth]{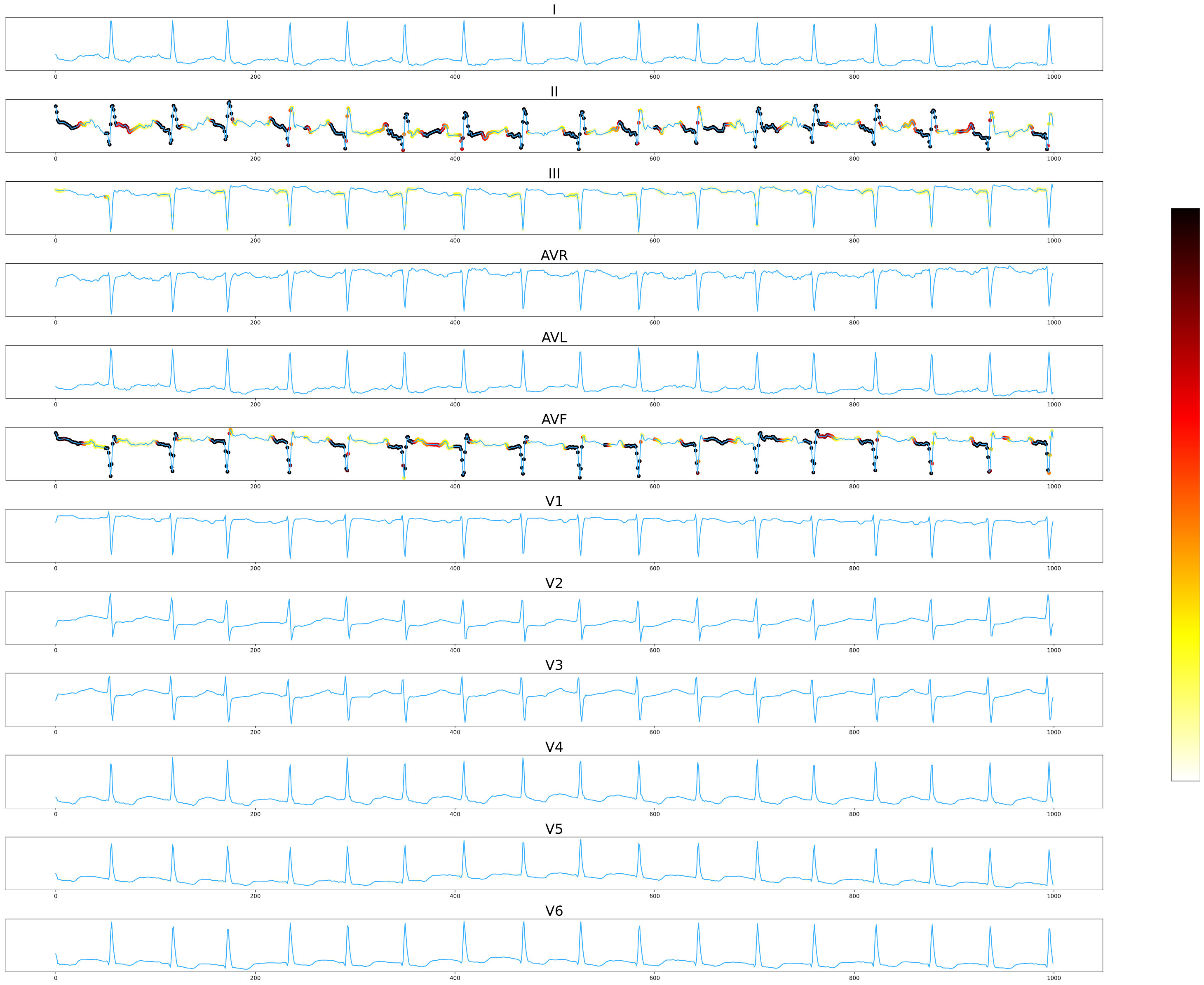}
  \caption{Visualization of normalized attention scores with red having a higher attention score and yellow having a lower attention score }
  \label{Fig:Fig4}
  \hfill
  \hfill
  \newline
  \centering
  \includegraphics[width=0.43\textwidth]{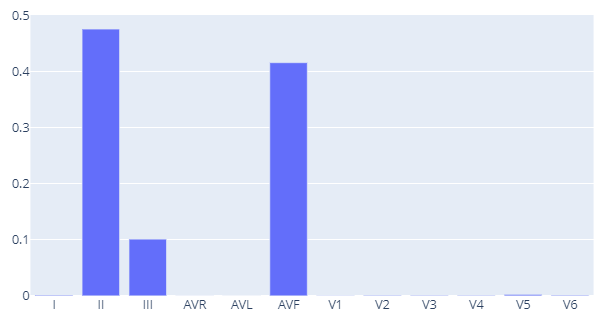}
  \caption{Channel Importance scores}
  \label{Fig:Fig5}%

\end{figure*}

\begin{table}[!h]
\setlength{\extrarowheight}{2pt}
\begin{center}
\caption{Class-wise ROC-AUC comparison}
\begin{tabular}{ | m{8em} || m{0.7cm}| m{0.7cm} | m{0.7cm} | m{0.8cm} | m{0.8cm} |  } 
\hline
 & CD & HYP & MI & NORM & STTC \\ 
\hline
Resnet101~\cite{He_2016_CVPR}  & 0.8928 & 0.8636 & 0.8921 &0.9098 & 0.9178\\ 
\hline
Mousavi et al.~\cite{8683140}  & 0.8555 & 0.8093 & 0.8708 &0.9096 & 0.8817\\
\hline
ECGNet~\cite{8438739}  & 0.9102 & 0.8842 & 0.9021 & 0.9336& 0.9205\\
\hline
Rajpurkar et al.~\cite{rajpurkar} & 0.8910 & \textbf{0.9185} & 0.9066 &0.9309&0.9306 \\
\hline
\textbf{IMLE-Net} & \textbf{0.9196} & 0.8834 & \textbf{0.9244} &\textbf{0.9450} & \textbf{0.9354}\\
\hline
\end{tabular}
\label{table:1}
\end{center}
\end{table}

\begin{table}[!h]
\setlength{\extrarowheight}{2pt}
\begin{center}
\caption{Class-wise accuracy comparison}
\begin{tabular}{ | m{8em} || m{0.7cm}| m{0.7cm} | m{0.7cm} | m{0.8cm} | m{0.8cm} |} 
\hline
 & CD & HYP & MI & NORM & STTC\\ 
\hline 

Resnet101~\cite{He_2016_CVPR}  & 87.74 & 90.89 & 83.12 &84.69 & 87.47 \\ 
\hline
Mousavi et al.~\cite{8683140} & 86.26 & 87.19 & 82.47 &81.78 & 83.26\\
\hline
ECGNet~\cite{8438739} & 88.71 & 90.66 & 84.60 & 85.89 & 86.87 \\
\hline
Rajpurkar et al.~\cite{rajpurkar} & \textbf{90.01} & 88.76 & 85.94 & 85.94 & 88.90 \\
\hline
\textbf{IMLE-Net} & 88.53 & \textbf{91.58} & \textbf{87.97} &\textbf{87.10} & \textbf{89.08}\\
\hline
\end{tabular}
\label{table:2}
\end{center}
\end{table}

\begin{table}[!h]

\setlength{\extrarowheight}{2pt}
\begin{center}
\caption{Overall Performance comparison}
\begin{tabular}{ | m{8em} || m{5em}| m{6em} | m{5em} | } 
\hline
& Macro ROC-AUC& Mean Accuracy & F1 score(Max)\\ 
\hline
Resnet101~\cite{He_2016_CVPR} & 0.8952 & 86.78 & 0.7558 \\ 
\hline
Mousavi et al.~\cite{8683140} &0.8654 & 84.19 & 0.7315 \\
\hline
ECGNet~\cite{8438739} &0.9101 & 87.35 & 0.7712 \\
\hline
Rajpurkar et al.~\cite{rajpurkar} &0.9155 & 87.91 & 0.7895  \\
\hline
\textbf{IMLE-Net} & \textbf{0.9216} & \textbf{88.85} & \textbf{0.8057}\\
\hline
\end{tabular}
\label{table:3}
\end{center}
\end{table}

\begin{table}[!h]
\setlength{\extrarowheight}{2pt}
\begin{center}
\caption{Accuracy and ROC-AUC metrics for two subtypes of Myocardial Infarction}
\begin{tabular}{ |m{6em}|| m{4 em }|m{4em} | m{4em}|m{4em}|}
\hline
& ASMI & IMI & NORM & Overall \\
 \hline
 Accuracy  & 89.47  & 87.24& 94.09 & 90.27 \\
 \hline
 ROC-AUC &  0.9594 & 0.9336  &0.9906 &0.9612  \\
 \hline
\end{tabular}
\label{tab:table-4}
\end{center}
\end{table}

The performance metrics for the selected subtypes of MI are shown in~\Cref{tab:table-4}. The metrics shown in this table are class-wise accuracy and class-wise ROC-AUC scores along with mean accuracy and macro ROC-AUC score. The best performance is observed for the NORM class followed by the ASMI and IMI classes.

The beat level attention scores, the rhythm level attention scores and the channel level attention scores are visualized for an ECG recording in \Cref{Fig:Fig4} with \Cref{Fig:Fig5} describing the importance of a specific channel responsible for the classification. The patient in the ECG recording shown in \Cref{Fig:Fig4} has an inferior myocardial infarction (IMI) present in leads II, III and aVF. Our proposed model was able to identify the channels responsible for this particular subtype of MI in~\Cref{Fig:Fig5}. These findings have been validated by an independent clinician. The channel level attention scores assess the relevance of specific channels in an ECG recording. This setting is crucial as the abnormal beat characteristics might be spread only over a few of the channels and may be completely absent in the rest of the channels. So along with the beat level, rhythm level interpretations, it is also essential to capture such channel level interpretations.

\begin{table}[!h]
\setlength{\extrarowheight}{2pt}
\begin{center}

\caption {Accuracy score comparison on different ECG lead configurations (Ablation Study)}
\begin{tabular}{ |m{5em}|| m{2em}|m{2em}|m{2em}|m{3em}|m{3em}|m{3em}|}
\hline
 & CD & HYP & MI & NORM & STTC & Mean Accuracy\\
 \hline
 Limb Leads   & 87.97  & 89.59& 83.63 & 84.48 &85.85 & 86.38\\
 \hline
 Precordial Leads &  87.47 & 90.98  &84.46 &84.97 &88.48 & 87.27\\
 \hline
 Leads I,II,III &87.51 &88.95&83.40&85.25 &85.85 &86.19\\
 \hline
 12 Leads & \textbf{88.53} & \textbf{91.58}& \textbf{87.97} & \textbf{87.10}&\textbf{89.08} & \textbf{88.85}\\
 \hline
\end{tabular}
\label{table:5}
\end{center}
\end{table}

\begin{table}[!h]
\setlength{\extrarowheight}{2pt}
\begin{center}
\caption{ROC-AUC score comparison on different ECG lead configurations (Ablation Study)}
\begin{tabular}{ |m{4em}|| m{2.5em}|m{2.5em}|m{2.5em}|m{3em}|m{3em}|m{2.5em}|}
\hline
  & CD & HYP & MI & NORM & STTC & Macro ROC-AUC\\
 \hline
 Limb Leads   & 0.8725  & 0.8407& 0.8787 & 0.9292 &0.9120 & 0.8866\\
 \hline
 Precordial Leads &  0.8972 & 0.8627  &0.8683 &0.9290 &0.9261 &0.8966\\
 \hline
 Leads I,II,III &0.8755 &0.8447&0.8609&0.9259 &0.9106 & 0.8836\\
 \hline
 12 Leads & \textbf{0.9196} & \textbf{0.8834}& \textbf{0.9244} & \textbf{0.9450}&\textbf{0.9354} & \textbf{0.9216}\\
 \hline
\end{tabular}
\label{table:6}
\end{center}
\end{table}

Tables \ref{table:5} and \ref{table:6} report the class-wise accuracy and class-wise ROC-AUC scores for several different channel combinations as part of the ablation studies. It can be seen that the performance metrics when utilizing all the available 12 channels in the ECG recording are the highest for each class and overall when compared against different combinations of channels. The precordial leads gave the second-best overall performance when compared against limb leads and leads I, II and III. Further, disorder classes such as HYP and STTC were classified with better metrics. Limb leads and leads I, II and III gave similar performance metrics with marginally better performance on the limb leads. In short, the standard 12-lead ECG setting gives the best performance and the performance drops as the leads are removed due to loss of information in the channels, again emphasizing the importance of multi-channel information in diagnosing heart conditions.

\section{Conclusion}
In this study, we proposed an interpretable multi-level model for multi-channel ECG classification. The proposed model has been evaluated on the PTB-XL dataset and it outperforms several of the existing models. An ablation study was performed to understand the changes in the performance of the model with different combinations of ECG channels. The attention scores generated were visualized and compared with guidelines that cardiologists follow to diagnose two subtypes of myocardial infarction. The interpretability aspect of the proposed model enables a better understanding of how it can be useful in clinical decision-making in a real-life scenario. In future, the model interpretability can be evaluated against a wide range of heart pathologies.

\section*{Acknowledgement}

We thank IHub-Data, IIIT Hyderabad for financial support and Dr. Adithi Devi for validating attention visualizations.

\printbibliography

\addtolength{\textheight}{-12cm}   




\end{document}